\documentclass[fleqn,usenatbib]{mnras}

\usepackage{newtxtext,newtxmath}

\usepackage[T1]{fontenc}

\DeclareRobustCommand{\VAN}[3]{#2}
\let\VANthebibliography\thebibliography
\def\thebibliography{\DeclareRobustCommand{\VAN}[3]{##3}\VANthebibliography}

\usepackage{graphicx}	
\usepackage{amsmath}	

\newcommand{\new}{\color{black}}

\title[Precise Diffusion Coefficients for White Dwarfs]{Precise Diffusion Coefficients for White Dwarf Astrophysics}

\author[M. E. Caplan \& I. F. Freeman]{
M. E. Caplan,$^{1}$\thanks{E-mail: mecapl1@ilstu.edu}
I. F. Freeman,$^{1}$
\\
$^{1}$Department of Physics, Illinois State University, Normal, IL 67190, USA\\
}

\date{Accepted XXX. Received YYY; in original form ZZZ}

\pubyear{2021}

\begin{document}
\label{firstpage}
\pagerange{\pageref{firstpage}--\pageref{lastpage}}
\maketitle

\begin{abstract}
Observations of galactic white dwarfs with Gaia have allowed for unprecedented modeling of white dwarf cooling, resolving core crystallization and sedimentary heating from neutron rich nuclei. These cooling sequences are sensitive to the diffusion coefficients of nuclei in Coulomb plasmas which have order 10\% uncertainty and are often not valid across coupling regimes. Using large scale molecular dynamics simulations we calculate diffusion coefficients at high resolution in the regime relevant for white dwarf modeling. We present a physically motivated law for diffusion with a semi-empirical correction which is accurate at the percent level. Implemented along with linear mixing in stellar evolution codes, this law should reduce the error from diffusion coefficients by an order of magnitude.
\end{abstract}




\section{Introduction}

The Gaia DR2 has allowed modelers to resolve a number of phenomena in white dwarf (WD) astrophysics with unprecedented precision. The latent heat release from core crystallization, first theorized by \cite{van1968crystallization}, has now been confirmed \citep{Winget_2009,tremblay2019core}. Furthermore, the sedimentary heating of neutron rich nuclei, especially $^{22}$Ne \citep{isern1991role,bildsten2001gravitational}, has also been resolved and may explain the anomalous heat source in the so-called `Q branch' WDs \citep{cheng2019cooling}. These WDs may be the products of double WD mergers, which could produce the inferred excess of $^{22}$Ne \citep{Cheng_2019_IAU,Camisassa2020}. These recent successes have driven theoretical efforts to improve inputs to those models; for example, a new C/O phase diagram has been reported by \citealt{blouin2020toward} and crystallization of C/O/Ne (and C/O/Fe) mixtures has been revisited with MD \citep{caplan2020neon}. Notably, the abundance of $^{22}$Ne inferred by modelers is ultimately dependent on diffusion coefficients $D$ which set cooling timescales \citep{bildsten2001gravitational,bauer2020multi}.
Furthermore, a recent Letter has argued that actinide separation in the core may provide a heat source to trigger a type 1a supernova in sub-Chandrasekhar WDs; this mechanism is sensitive to the diffusion coefficient of $U$ \citep{HorowitzCaplanPRL2021}. In summary, we are entering an era of WD astrophysics where microphysics must be accurate at the percent level for modeling, which motivates us to revisit diffusion in WDs.

Both \citet{Camisassa2020} and \cite{bauer2020multi} use $D$ from \citet{hughto2010diffusion}, which is generally representative of the literature computing diffusion rates in WDs. In similar works, $D$ is found at a few temperatures (or densities) from molecular dynamics (MD) simulations but without clear uncertainties. The reported fits match the MD to about 10\% and can have order of magnitude errors when extrapolated outside the parameter range studied by the MD. For this new era of precision white dwarf modeling, the problem of nuclear diffusion in WDs must be revisited.

In WD interiors nuclei are effectively point particles of charge $eZ$ immersed in a neutralizing background gas of degenerate electrons, which screens a Coulomb repulsion, 

\begin{equation}\label{eq.V}
    V_{ij}(r) = \frac{e^2 Z_i Z_j}{r_{ij}} e^{-r_{ij}/\lambda}
\end{equation}

\noindent with inter-particle separation $r_{ij}$ and Thomas-Fermi screening length $\lambda^{-1} = 2 k_F \sqrt{\alpha/\pi}$ (electron Fermi momentum $k_F = (3 \pi^2 \langle Z \rangle n)^{1/3}$, ion number density $n$). These `Coulomb plasmas' are widely studied for their applications in astrophysics, dusty plasmas, and terrestrial fusion experiments. The one-component plasma (OCP) is characterized by two parameters: the coupling constant $\Gamma = e^2 Z^2 / a T$ (Wigner Seitz radius $a=(3/4 \pi n)^{1/3}$, temperature $T$), and the screening length $\lambda$. The dimensionless structure factor $\kappa^{-1} = \lambda / a $ is often used in place of $\lambda$ \citep{PhysRevE.66.016404,hughto2010diffusion}. At higher order the charge-to-mass ratio of the nucleus $Z/A$ may be important; this is typically neglected in WD astrophysics as most nuclei are the result of alpha-burning ($Z/A=0.5$).
Above (below) the critical $\Gamma_{\mathrm{crit}} = 175$ the OCP is solid (liquid) \citep{potekhin2000equation}.

Considerable progress studying Coulomb plasmas has been made in the past two decades (see \citealt{RevModPhys.81.1353,michaud2015atomic,CaplanRMP} for reviews). Diffusion coefficients in particular have been studied for binary mixtures and across coupling regimes; in this work, we seek to apply to WDs the works of \cite{PhysRevE.71.036408,PhysRevLett.108.225004,khrapak2013,PhysRevE.95.013206}.

In this work, we develop a theoretically motivated law for the diffusion coefficients of the OCP in WDs which is valid across coupling regimes (Sec. \ref{sec:theory}) and fit to new high resolution MD simulations (Secs. \ref{sec:md} and \ref{sec:d}). Our law requires a semi-empirical correction which makes our form of the diffusion coefficient accurate to 1\%.

\section{Modeling Diffusion Across Coupling Regimes}\label{sec:theory}

Our model follows straightforwardly from \cite{PhysRevLett.108.225004}, which used piecewise functions to model diffusion smoothly across coupling regimes, though primarily for mixtures relevant to inertial-confinement fusion. Here we will attempt to consolidate this treatment into a single analytic expression for $D$. We adopt the normalization convention $D^* = D/\omega_p a^2$, with ion plasma frequency $\omega_p = ( 4 \pi e^2 \langle Z \rangle^2 n / \langle M \rangle )^{1/2}$ and Wigner-Seitz radius $a$.

Chapman-Spitzer (CS) theory gives an analytic expression for the Coulomb logarithm $ \ln(\Lambda_{CS}) = \ln(C \Gamma^{-3/2})$ from collision theory in the weak coupling limit, $\Gamma \lesssim 1$, which causes $D^*_{CS} \propto \Gamma^{-5/2}/\ln(\Lambda_{CS})$ to diverge to negative infinity in the limit of $\Gamma \to 1$ from below. Both \cite{PhysRevLett.108.225004} and \cite{khrapak2013} extend Chapman-Spitzer into the strongly coupled regime with an approximate form of the Coulomb logarithm, to $\ln(\Lambda_{CS})=\ln(1 + C \Gamma^{-3/2})$, a well known trick. This modification has the desired power-law behavior through the moderately to strongly coupled regime (known since \citealt{Hansen1975}). This extended CS model will form the basis of our model here,

\begin{equation}\label{eq:DECS}
    D^*_{CS}(\Gamma) = \sqrt{\frac{\pi}{3}}  \frac{\Gamma^{-5/2}}{ \ln(1 + C \Gamma^{-3/2})}
\end{equation}

\noindent where the free parameter $C$ may depend on $\lambda$ \citep{PhysRevLett.108.225004,khrapak2013}. \cite{khrapak2013} reports that this extension approximately matches the MD to about 30\% using $C=0.4$; their Figs. 1 and 3 show that it underpredicts in the moderately coupled regime.
As we are primarily interested in $\Gamma \gtrsim 1$, it is insightful to consider the series expansion at large $\Gamma$,

\begin{equation}
\sqrt{ \frac{\pi}{3}} \frac{\Gamma^{-5/2}}{\ln(1+C\Gamma^{-3/2} ) } \approx  \sqrt{ \frac{\pi}{3}} \left[ \frac{\Gamma^{-1}}{C } +  \frac{\Gamma^{-5/2}}{2 } + ... \right]
\end{equation}

\noindent so we are effectively using a $\Gamma^{-1}$ power-law. 

In the strongly coupled regime ($\Gamma \gg 1$) ions become trapped and diffusion proceeds via thermally activated jumps between `cages' of its neighbors, which is well described by an Eyring model, $D^* = (A/\Gamma) e^{-B \Gamma}$ \citep{hughto2010diffusion,PhysRevLett.108.225004}\footnote{Variations using $\Gamma^\alpha$ are also used occasionally, see \cite{hughto2010diffusion}.}. We observe the limiting behavior of Eq. \ref{eq:DECS} is  $D_{CS}^* \approx \sqrt(\pi/3)/(C\Gamma)$ for $\Gamma \gtrsim 1$ from our expansion, which resembles the prefactor in an Eyring model with $A = \sqrt{(\pi/3)}/C$. Since the exponential term is suppressed at $\Gamma \ll \Gamma_{\mathrm{crit}}$ we can include it in the extended CS without effecting its performance in the weakly coupled regime. Therefore, we use the model

\begin{equation}\label{eq:DECS2}
    D^*(\Gamma) = \sqrt{\frac{\pi}{3}}  \frac{\Gamma^{-5/2}}{ \ln(1 + C \Gamma^{-3/2})} e^{-B \Gamma}
\end{equation}
\noindent with $B \sim \Gamma_{\mathrm{crit}}^{-1}$ and $C \sim  1$. $D^*$ in Eq. \ref{eq:DECS2} has limiting behavior of both the CS and Eyring models and should be straightforward to implement in WD modeling codes.  

In the following sections we will describe our procedure for calculating $D^*$ from MD simulations and fits for $B$ and $C$, as well as our empirical correction for higher order effects.

\section{Molecular Dynamics Formalism}\label{sec:md}

Our MD formalism has been used extensively to study Coulomb plasmas, see any of \cite{CaplanRMP,Caplan2018,Caplan2020,caplan2020neon} for details. Briefly, we simulate using the Indiana University Molecular Dynamics CUDA-Fortran code v6.3.1. We evolve an N-body system using the velocity Verlet algorithm with the potential from Eq. \ref{eq.V} computed between all pairs of particles to the nearest periodic image in a cubic simulation volume, with no cut-off distance for the potential. \new{Every particle is therefore treated as if it were in the center of a cube with a size equivalent to the simulation volume.}

Past work has generally used the velocity auto-correlation function (VACF) to determine $D$. This method requires large amounts of simulation data spaced closely in time in order to determine $\langle  \mathbf{v}_j (t_0 + t) \cdot \mathbf{v}_j(t_0)  \rangle$ \citep{hughto2010diffusion}. In contrast, we use the mean square displacement (MSD) in this work, which only requires a few configurations well spaced in time to determine displacements. These approaches are effectively equivalent from the perspective of the Green-Kubo relations, with $D$ being an integral of the VACF and the MSD being a derivative. 

The MSD has subtleties. To start, relatively large  simulation sizes must be used, requiring large particle number $N$, so that nuclei do not become fully diffused across the periodic boundary. It also requires that we determine the MSD twice; at very early time  $\langle x^2 \rangle \propto t^2$ as motion is ballistic, and only after a few collisions is $\langle x^2 \rangle \propto t$. Therefore, we compute from our simulations 

\begin{equation}
    D^* = \frac{  \langle \left| \mathbf{x}(2\Delta t) -  \mathbf{x}(0) \right|^2 \rangle - \langle \left| \mathbf{x}(\Delta t) -  \mathbf{x}(0) \right|^2 \rangle }{6\Delta t \omega_p a^2} 
\end{equation}
\noindent where $\Delta t$ is about $25 \omega_p^{-1}$ (approx. 500 MD timsteps).

Our greater accuracy is made possible by advances in high performance computing on GPUs, allowing us to run larger simulations for longer; our simulations use $N= 65536$ nuclei ($2^{16}$, optimized for parallelism). 
These large simulations have a box-side length $L \approx 65 a$ and resolve $D^*$ for $2.5 \lesssim \Gamma \lesssim 350$. In the weakly coupled regime nuclei become ballistic and fully diffuse in the simulation volume too quickly for MD to resolve, 
while the larger thermal velocities require smaller timesteps to be treated accurately. At high $\Gamma$, we are limited by supercooling of the material and exponential suppression of diffusive jumps. Checks for finite size effects with larger and smaller simulations, and using larger steps for temperature rescaling gave consistent $D^*$ in the relevant range of $\Gamma$.

Individual simulations were not confined to a single $\Gamma$. By periodically rescaling the particle velocities to a Maxwell Boltzmann distribution with a temperature smaller by some $\Delta T$ a single simulation can systematically step through $\Gamma$. The MSD for each interval in $T$ gives us $D^*$ at some $\Gamma$, thus determining $D^*$ to high precision and high resolution in $\Gamma$ with relative ease. Each interval simulates quickly, requiring only a thousand MD timesteps; it is the large particle number that ultimately gives us good statistics for diffusion. Over a given time interval at fixed $\Gamma$ the distribution of the MSD is approximately Poissonian, as expected. This scheme gives us points equally spaced in $\Delta T \propto \Gamma^{-1}$, clustering our data at low $\Gamma$. As such, we combine two sets of simulations which separately resolve $\Gamma \lesssim 35$ and $\Gamma \gtrsim 35$. We report on MD simulations at 12 points in our parameter space: at $\kappa^{-1} =$ 2.361 (Mg), 2.509 (Ne), 2.703 (O), and 2.975 (C) at $Z/A =$ 0.4, 0.45, and 0.5. For each, $D^*$ is calculated at about 300 values of $\Gamma$. 
Exact diffusion coefficients computed from our MD are available in the supplemental materials.

\begin{figure*}
\centering
\begin{minipage}{.47\textwidth}
\includegraphics[trim=10 120 155 10,clip,width=1.0\textwidth]{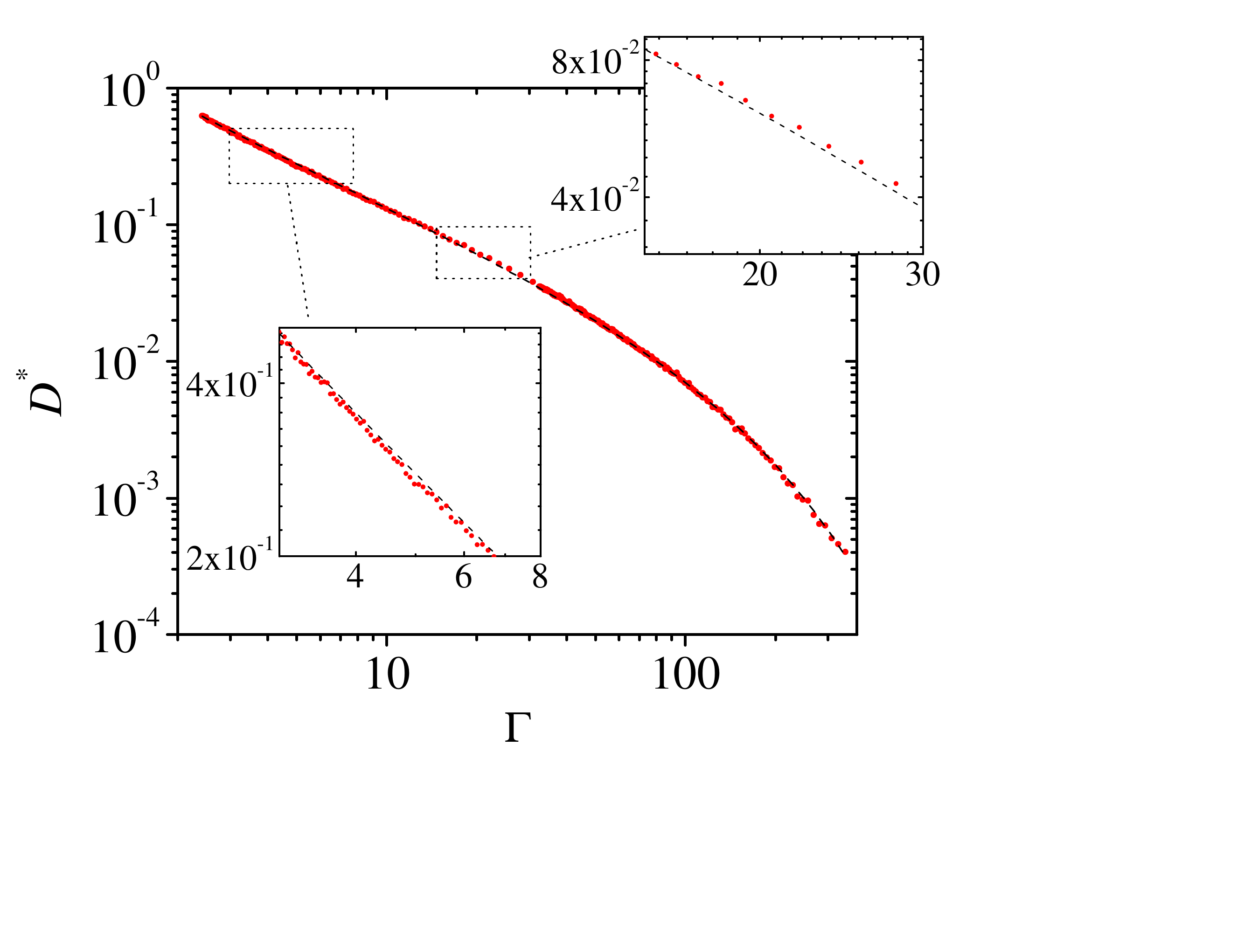}
\caption{\label{fig:Dstar} $D^*$ for $^{16}$O from MD (red) with  fit from Eq. \ref{eq:DECS2} (dashes) using best fit parameters $C=0.7323
 $ and $B=0.006937$. The fit is good to about 5\% for all $\Gamma$, but it is clear that it systematically overpredicts for $\Gamma \lesssim 10$ (bottom inset) and underpredicts for $10 \lesssim \Gamma \lesssim 100$ (top inset). Normalized residuals for all runs are shown in Fig. \ref{fig:resnocorrection}. }	
 \end{minipage}
 \hspace{.05\linewidth}
\centering
\begin{minipage}{.47\textwidth}
\includegraphics[trim=25 140 175 20,clip,width=1.0\textwidth]{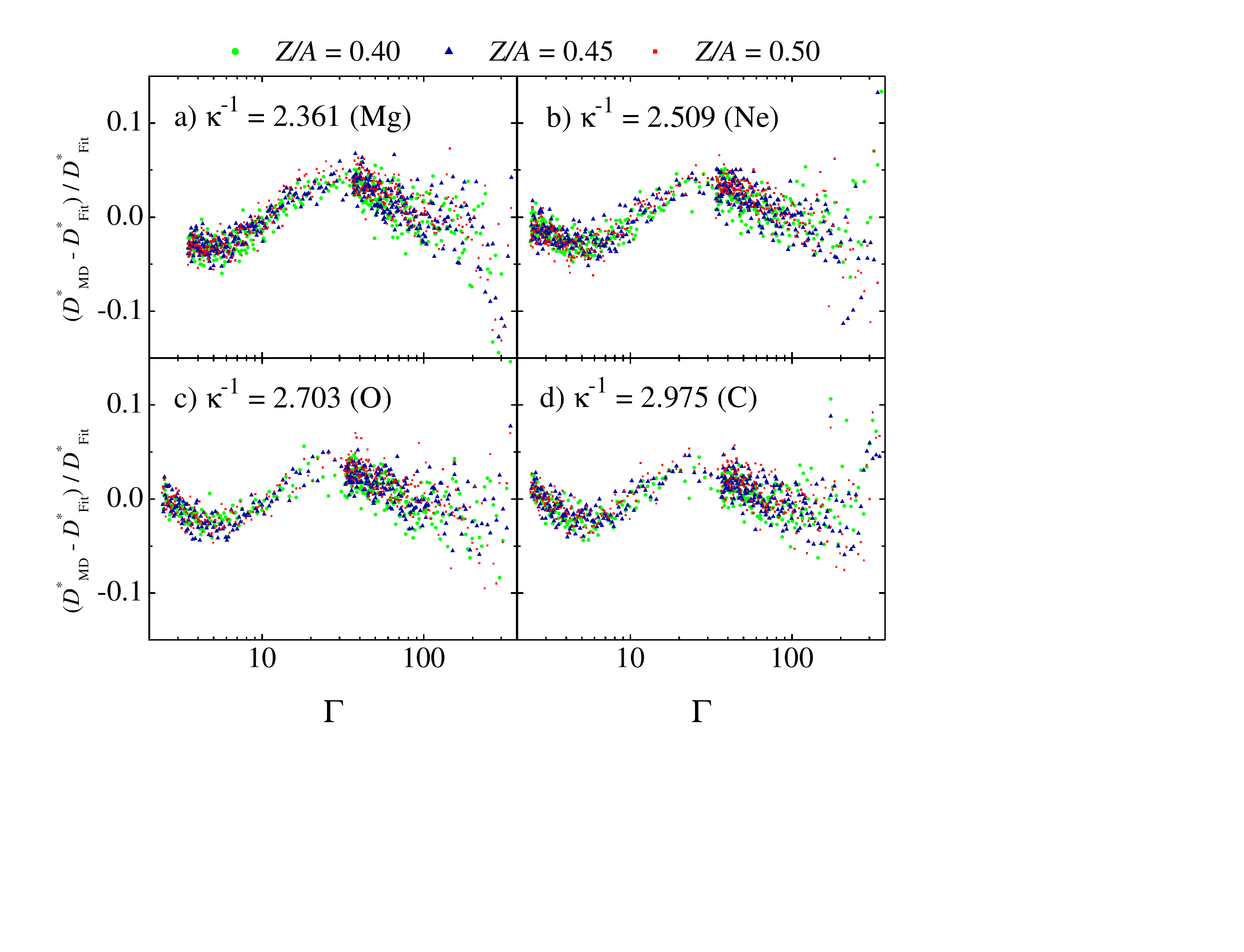}
\caption{\label{fig:resnocorrection} Normalized residuals between the MD and best fits to Eq. \ref{eq:DECS2} show approximately 5\% systematic uncertainty depending on $\kappa^{-1}$. Normalized diffusion coefficients are independent of the charge to mass ratio at the resolution our MD can resolve.}	
 \end{minipage}
\end{figure*}

\section{Diffusion Coefficients from Molecular Dynamics }\label{sec:d}

In Fig. \ref{fig:Dstar} we show $D^*(\Gamma)$ for $^{16}$O, along with our Eq. \ref{eq:DECS2}, which is representative of all our MD cases. The best fit parameters for Eq. \ref{eq:DECS2} over the twelve data sets give us $C=0.0333\kappa^{-1} + 0.6423$ and $B=0.006937$. The fractional residuals for all twelve cases are shown in Fig. \ref{fig:resnocorrection}. It is clear that our model very slightly underpredicts for $\Gamma \lesssim 10 $ and overpredicts for $10 \lesssim \Gamma \lesssim 100$; all our simulations have roughly 5\% systematic variation from Eq. \ref{eq:DECS2} with approximately 1\% variation from thermal noise in the MD. We do not resolve any dependence of $D^*$ on $Z/A$ (though the reader should recall $D = D^* \omega_p a^2$ where $\omega_p$ varies between isotopes). 

This residual is known in the literature and is not an artifact of our model or our scheme for computing $D^*$; in the inset of Fig. 1 of \cite{PhysRevLett.108.225004} a similar feature is weakly resolved. This feature has been difficult to model from first principles; for example, \citealt{PhysRevE.95.013206} computes corrections from effective potential theory for the one component plasma and binary mixtures, and are generally good at the 10\% level, but can have errors as large as 20-40\%. As such, we report an empirical correction for this residual.

The deviation of our MD from the Eyring model may be considered evidence for non-Arrenhius effects, which form the basis of our correction. The peaks observed in the residual are approximately normally distributed in $\ln(\Gamma)$ so we conjecture that this may be due to many-body correlations.  
Eyring models generally take `hops' between cages of neighbors to have a constant activation energy and treat diffusion effectively as a one-body Poisson process. However, diffusion requires that a `site' must be vacated before a hop can proceed, suggesting that the existence of holes and many-body correlations may be important such that the activation energy has a more complicated temperature dependence at higher order. For example, two-body correlations such as exchanges may occur, while three-body (or greater) correlations may involve small loops of exchanges or cascades as a vacancy forms and diffuses a short distance, which could be studied in more detail in future work. If a one-body jump is an independent process in $\Gamma$, then a product of many such processes would result in a lognormally distributed many-body diffusion process, and could explain the deviation from the simple Eyring model in the strongly coupled regime. 
While non-Arrhenius diffusion has now been resolved in other systems \citep{PhysRevLett.112.145901,PhysRevB.102.184110}, such models are less frequently applied to Coulomb plasmas.

We propose adding a term to our model for many-body correlations, 

\begin{equation}
    D^* = D^*_1 + D^*_{2+}
\end{equation}

\noindent with $D^*_1$ for one-body diffusion (effectively our Eq. \ref{eq:DECS2}) and $D^*_{2+}$ a correction for many-body diffusion, where

\begin{equation}\label{eq:DECSEM2}
\begin{aligned}
    D^*_{2+}  =   \sqrt{\frac{\pi}{3}}   \frac{\Gamma^{-5/2}}{ \ln(1 + C \Gamma^{-3/2})}  \epsilon \left( -e^{-B \Gamma - \ln( \Gamma/\Gamma_1)^2} 
      +\   e^{-B \Gamma  - \ln( \Gamma/\Gamma_2)^2}  \right)
\end{aligned}
\end{equation}

\noindent as in Eyring-Lognormal distributions \citep{modarres2017probabilistic}. Each exponential produces a single bell-shaped curve in $\ln{(\Gamma)}$ roughly centered at $\Gamma_1 (\approx 5)$ and $\Gamma_2 (\approx 30)$ with amplitude $\epsilon (\approx 0.05)$ so that the function has the qualitative shape of a single period of a sine curve to match the residual in Fig. \ref{fig:resnocorrection}. This correction converges quickly to zero outside the range of $1 \lesssim \Gamma \lesssim 100$ so they do not change the asymptotic behavior of our original model in Eq. \ref{eq:DECS2}. For simplicity, the added Eyring-Lognormal exponentials use the same prefactor as the one-body term, taken to be the extended CS and implicitly contain $e^{-B\Gamma}$. \footnote{We also considered a two-parameter function with a single lognormal term with $\Gamma_2 \approx 30$ and $\epsilon \approx 0.07$. This fit required greater $C$ and the systematic error in the residual remained apparent.} We again emphasize that while our correction term is physically motivated it should be considered empirical.

Taken together, our final model for diffusion is

\begin{equation}\label{eq:DECSfinal}
\begin{aligned}
    D^*(\Gamma)  =  \sqrt{\frac{\pi}{3}}   \frac{\Gamma^{-5/2}}{ \ln(1 + C \Gamma^{-3/2})}  \left( e^{-B \Gamma}\right) & \left( 1 -  \epsilon e^{ - \ln( \Gamma/\Gamma_1)^2} \right.\\  
     & \left. +\  \epsilon e^{ - \ln( \Gamma/\Gamma_2)^2}  \right)
\end{aligned}
\end{equation}

\noindent with best fit parameters

\begin{equation}\label{eq:params}
\begin{aligned}
    C & =  0.0333 \kappa^{-1} + 0.6423 \\
    B & =  0.006937 \\
    \epsilon & =  0.03372\kappa^{-1} -0.04373 \\
    \Gamma_1 & =  5.767\kappa^{-1} -8.331 \\
    \Gamma_2 & =  -17.057\kappa^{-1} + 65.282
\end{aligned}
\end{equation}

\begin{figure}
\centering
\begin{minipage}{.47\textwidth}
\includegraphics[trim=25 140 175 20,clip,width=1.0\textwidth]{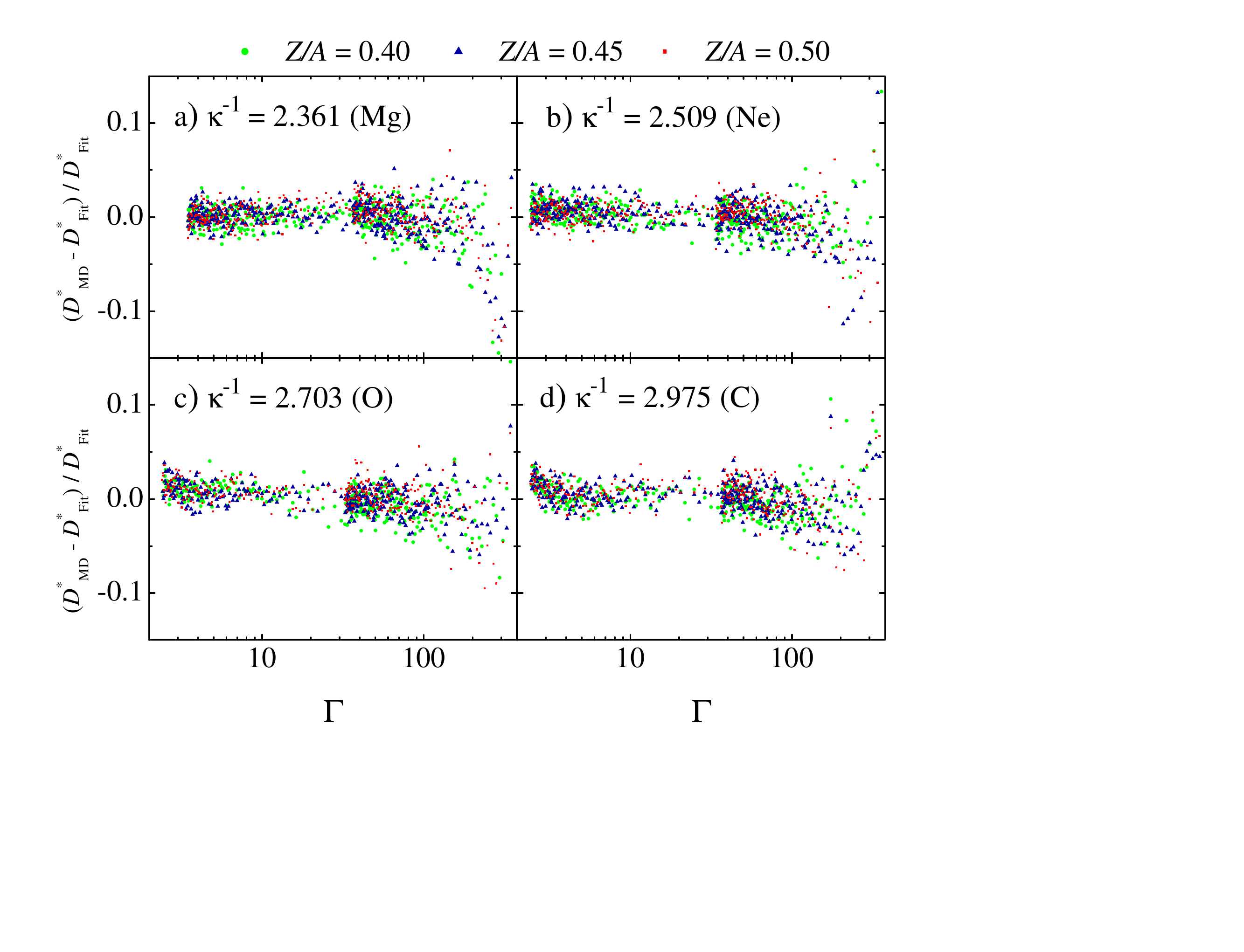}
\caption{\label{fig:rescorrection} Normalized residuals between the MD and best fits to Eq. \ref{eq:DECSfinal}. The 2\% thermal noise dominates over remaining systematic uncertainties.}
 \end{minipage}
\end{figure}

\noindent In Fig. \ref{fig:rescorrection} we show the residuals between the MD and Eq. \ref{eq:DECSfinal}. The error is largely due to thermal noise at the level of 2\%. At $\Gamma>200$ our model fits to about 10\%, but this is dominated by thermal noise in the supercooled MD rather than any systematic failure of the model. \new{At high $\Gamma$ diffusion proceeds very slowly and so the noise can be attributed to to poor statistics and a smaller effective sample size. Longer simulations do not necessarily improve the situation; at such high $\Gamma$ the MD configuration has a tendency to spontaneously crystallize as the supercooled state is metastable. Modelers interested in including the phase transition should look to work such as \cite{PhysRevE.84.016401} which calculates diffusion coefficients in Coulomb crystals and gives examples of $D^{*}$ defined piecewise to handle the discontinuity due to crystallization.}

\new{However, it is possible that the model does has some systematic difficulty at high $\Gamma$ ($\gtrsim 200$). In Fig. \ref{fig:rescorrection} it seems that the model may systematically underpredict for $\kappa^{-1} = 2.361$ at high $\Gamma$ but may overpredict at $\kappa^{-1}=2.975$. Again, it is unclear if this is due to a systematic effect in the model, or if this is just a coincidence due to the shape of the noise in the MD. As our MD results are available in the supplemental materials future authors are free to refit the parameters in Eq. \ref{eq:params}, perhaps quadratically in $\kappa^{-1}$. However, if there is some systematic difficulty at very high $\Gamma$ it may not be so important for astrophysics models where the mixture is dominated by low $Z$ nuclei which should not be expected to be a supercooled component in a mixture.}

\section{Discussion}

We are optimistic that Eq. \ref{eq:DECSfinal} and the parameters in Eq. \ref{eq:params} will be useful for WD astrophysics. The higher precision of this model will effectively eliminate the modeling uncertainty stemming from diffusion coefficients at the resolution needed for present and near-future work. \new{The models in Eq. \ref{eq:DECS2} and Eq. \ref{eq:DECSfinal} also have applications in related problems in accreting neutron star physics, such as crust crystallization and the nucleation and phase separation of high $Z$ snow in the ocean, though more MD is required to fit the free parameters to an appropriate range of screening lengths and coupling factors which, which could be considered in future work.}

Our model is an improvement over \cite{hughto2010diffusion} in the regime most relevant for WD modeling, about $1 < \Gamma < \Gamma_{\mathrm{crit}}$, with uncertainty at the percent level which is largely due to the thermal noise in the MD. Our fit has the most trouble with $\Gamma > \Gamma_{\mathrm{crit}}$ with order 10\% error, however this may not be so important. Above $\Gamma_{\mathrm{crit}}$ mixtures tend to crystallize, so trace high $Z$ nuclei will not spend so long at such high $\Gamma$. 
The success of our model across coupling regimes should make it possible to model less abundant nuclides beyond $^{22}$Ne as heat sources, such as iron group nuclei, where $\Gamma_{\mathrm{Fe}} \approx 11 \Gamma_{\mathrm{C}}$. Future authors should be aware that while $D^*$ is not sensitive to $Z/A$ at the percent level, the dimensional diffusion coefficient $D = D^* \omega_p a^2$ depends on the nuclide mass through the plasma frequency, which must be treated properly when generalizing our model to mixtures and using the mixture averaged quantities.

The simplicity and success of our model for pure systems motivates future work generalizing it for mixtures of astrophysical interest. Linear mixing for binary mixtures is well established \citep{PhysRevLett.108.225004,PhysRevE.95.013206}, and it is likely straightforward to generalize the model in this work to astrophysically interesting ternary mixtures with a trace third component, such as Ne, Fe, or U. Since only these trace third components have a neutron excess, they are the only species for which accurate diffusion coefficients are needed. Thus, if the background can be accurately modeled as a binary mixture when neglecting the nuclides of interest, then it may be possible to reduce the problem of diffusion in ternary mixtures to a restricted binary mixture between a $Z/A=0.5$ background and that trace neutron rich nuclide, which will be verified with MD in future work.

\section*{Acknowledgements}

The authors thank C. J. Howoritz for conversation. This work benefited from support by the National Science Foundation under Grant No. PHY-1430152 (JINA Center for the Evolution of the Elements). The authors acknowledge the Indiana University Pervasive Technology Institute for providing supercomputing and database, storage resources that have contributed to the research results reported within this paper. This research was supported in part by Lilly Endowment, Inc., through its support for the Indiana University Pervasive Technology Institute. 

\section*{Data Availability}

The data underlying this article are available in the article and in its online supplementary material.



\bibliographystyle{mnras}
\bibliography{bib} 








\bsp	
\label{lastpage}
\end{document}